\documentclass[11pt]{article}

\begin{document}
\newcommand{\h}{{\mathcal H}}
\newcommand{\be}{\begin{eqnarray}}
\newcommand{\ee}{\end{eqnarray}}
\newcommand{\bes}{\begin{eqnarray*}}
\newcommand{\ees}{\end{eqnarray*}}

\title{On Minimum Uncertainty States}
\author{Pankaj Sharan\\{\small Physics Department, Jamia Millia Islamia, New Delhi, 110 025, India}}
\date{}
\maketitle
\begin{abstract} Necessary and sufficient condition for the existence of 
a minimum uncertainty state for an arbitrary pair of observables is given.
\end{abstract}

Let the states of a physical system be represented by normalized vectors in a Hilbert space
$\h$. For two vectors $\phi$ and $\psi$ in $\h$, denote the inner product by
$(\psi,\phi)$ and define the norm $\|\phi \|$ of $\phi$ by $\|\phi\|^2=(\phi,\phi)$.
Let $A$ and $B$ be two observables; that is, self-adjoint operators. Let the observable
$C$ be defined by the commutator $[A,B]=iC$.   The  expectation value
$(\psi,A\psi)$  of $A$ is denoted by $a$. Similarly, expectation values of 
$B$ and $C$ in the state $\psi$  are denoted $b$ and $c$ respectively. 

The statement of the uncertainty inequality is  
\be \Delta A \Delta B \geq \frac{1}{2}|c| ,\label{uncert}\ee
where the variance (or uncertainty) of $A$ in the state $\psi$ is defined as 
$\Delta A=\|(A-a)\psi\| $ and a similar formula for $\Delta B$.
We say that $\psi$ is a minimum uncertainty state (MUS) for the pair $A,B$ if the equality is 
achieved in (\ref{uncert}) above, that is, if
\be \Delta A \Delta B = \frac{1}{2}|c| .\ee

The proof of the uncertainty inequality is a direct application of the Schwarz 
inequality which states that 
\be |(\psi,\phi)| \leq \| \psi\| \|\phi\| \ee
 for any two vectors $\phi$ and $\psi$ in $\h$. 
We assume that one of the vectors (say $\phi$) is non-zero to avoid triviality.
The Schwarz inequality becomes an
equality if and only if $\psi$ can be written as the other (non-zero) vector $\phi$
multiplied by a complex number $z$
\be \psi=z \phi. \ee

The proof of the uncertainty inequality is as follows. Denote by Im $ z$ the imaginary part of a complex number $z$.  The Schwarz inequality implies
\bes \Delta A\Delta B &=& \|(A-a)\psi\| \|(B-b)\psi\| \\
&\geq & | ((A-a)\psi,(B-b)\psi)| \mbox{\hspace{1.6cm}               Ineqauality 1} \\
&\geq & | {\rm Im } ((A-a)\psi,(B-b)\psi)| \mbox{ \hspace{1cm}           Ineqauality 2} \\
&=& \left| \frac{1}{2i}\Big[((A-a)\psi,(B-b)\psi)-((B-b)\psi,(A-a)\psi)\Big]\right| \\
&=& \frac{1}{2}|c|.
\ees
The condition for $\psi$ to be a MUS for $A,B$ is that at both the places
above (Inequality 1 and 2) the equality must be satisfied. The first one is satisfied 
if and only if there is a complex number $z$ such that 
\be (A-a)\psi= z(B-b)\psi \label{a-ib}\ee
where we assume $\Delta B=\|(B-b)\psi\| \neq 0$ to avoid the trivial case when both
$\Delta A$ and $\Delta B$ are zero. By taking norm on both sides 
of the above equation we also note that 
\be \Delta A=|z| \Delta B .\label{modz}\ee

The second inequality (Inequality 2) becomes an equality if and only if the real part of 
$((A-a)\psi,(B-b)\psi)$ is zero. This happens if
\bes ((A-a)\psi,(B-b)\psi)+((B-b)\psi,(A-a)\psi)=0 \ees
which, in the light of $(A-a)\psi= z(B-b)\psi$ implies that Re $z=0$.  In other words,
$z=i\lambda$ for a real number $\lambda$. The magnitude of $\lambda$ 
follows from (\ref{modz}) above as
\be |\lambda|=\frac{\Delta A}{\Delta B} .\ee

To obtain the sign of $\lambda$ we proceed as follows. Write $z=i\lambda$ in (\ref{a-ib})
and calculate 
\be \|(A-i\lambda B)\psi\|^2 &= & |a-i\lambda b|^2=a^2+\lambda^2b^2. \label{sign}\ee
The left hand side is 
\bes \|(A-i\lambda B)\psi\|^2=((A-i\lambda B)\psi,(A-i\lambda B)\psi)
=(\psi,(A+i\lambda B)(A-i\lambda B)\psi), 
\ees
and 
\bes (A+i\lambda B)(A-i\lambda B)=A^2+\lambda^2B^2+\lambda C.
\ees
Substituting these in (\ref{sign}) and using $(\Delta A)^2=(\psi,A^2\psi)-a^2$, $\Delta A=|\lambda|\Delta B$
etc. we get,
\bes 2\lambda^2(\Delta B)^2+\lambda c=0\ees
which shows that the sign of $\lambda$ must be opposite to that of $c$.

With the notation as  above, we have proved the following theorem :

\begin{quote} 
{\em
For $\psi$ to be a MUS for the pair $A,B$ (with $\Delta B\neq 0$) the necessary and sufficient
condition is that 
\bes (A-a)\psi= i\lambda (B-b)\psi \ees
where $\lambda$ is a real number whose magnitude is given by 
$|\lambda|=\Delta A/\Delta B$ and whose sign is opposite to that of $c$.
}
\end{quote}

We see that the condition for MUS can also be written as
\be (A-i\lambda B)\psi=(a-i\lambda b)\psi ,\ee
which means that $\psi$ must be an eigenvector of the non-hermitian operator
$A-i\lambda B$ with the complex eigenvalue $a-i\lambda b$.

A well-known example of 
MUS is the gaussian wave-packets in one dimension:
\be \psi=\frac{1}{(2\pi \sigma^2)^{1/4}}\exp\left[ikx-\frac{(x-x_0)^2}{4\sigma^2}\right] \ee 
for the pair of operators $A=x$ and $B=-id/dx$. Here $a=x_0, b=k, \Delta A=\sigma$
and $\Delta B=1/(2\sigma)$. Thus $|\lambda|=2\sigma^2$, and because $c=1>0$ 
we have $\lambda=-2\sigma^2$. One can check that the wave packet above is the eigenfunction
of the operator  
\bes \left(x+2\sigma^2\frac{d}{dx}\right) \ees
with complex eigenvalue $x_0+2i\sigma^2 k$. 

\begin{center}{\bf Acknowledgement}\end{center}

I thank Pravabati Chingangbam and Tabish Qureshi for useful discussions
on the uncertainty inequality in 1995 and 2011 respectively. 
\end{document}